\documentclass[aps,prd,twocolumn,superscriptaddress,showpacs,showkeys,nofootinbib]{revtex4-1}
\usepackage[dvips]{graphicx}
\usepackage[english]{babel}
\usepackage{xcolor}
\usepackage{mathtools}
\usepackage{pdfsync}

\usepackage{psfrag}

\usepackage{amssymb,epsf,epsfig}
\usepackage{amsmath}

\usepackage{graphicx}

\usepackage{slashed}
\usepackage[active]{srcltx}

\def\d{\hbox{{d}\kern-.20em\hbox{l}}}

\def \matrix #1 {\left(\begin{array}{cc} #1 \end{array}\right)}

\def \tr {\mathop{\rm tr}\nolimits}

\def\II{\hbox{{1}\kern-.25em\hbox{l}}}

\newcommand \vev [1] {\langle{#1}\rangle}

\begin{document}


\title{
\hspace{15cm}
\phantom{ On  operator mixing  in fermionic CFTs in non-inte }
{ \textmd{DESY 18-149}}
\\[3mm]
On operator mixing  in fermionic CFTs in non-integer dimensions}

\date{\today}

   \author{Yao Ji}
   \email{yao.ji@ur.de}
\affiliation{Institut f\"ur Theoretische Physik, Universit\"at
   Regensburg, D-93040 Regensburg, Germany}

  \author{Alexander N. Manashov}
   \email{alexander.manashov@desy.de}
\affiliation{ Institut f\"ur Theoretische Physik, Universit\"at Hamburg
   D-22761 Hamburg, Germany}
   \affiliation{Institut f\"ur Theoretische Physik, Universit\"at
   Regensburg, D-93040 Regensburg, Germany}
\affiliation{
   St.Petersburg Department of Steklov
Mathematical Institute, 191023 St.Petersburg, Russia}

   \date{\today}

\begin{abstract}
We consider the renormalization of four-fermion operators in the critical QED and $SU(N_c)$ version  of Gross--Neveu--Yukawa
model in non-integer dimensions. Since the number of mixing operators is infinite, the diagonalization of an anomalous
dimension matrix becomes a nontrivial problem. At leading order, the construction of eigen-operators  is equivalent to solving
 certain three-term recurrence relations. We find analytic solutions of these recurrence relations that allow us to determine
 the spectrum of anomalous dimensions and study their properties.
\end{abstract}



\maketitle

\section{Introduction}
Quantum field theories (QFTs) in non-integer dimensions $d<4$ were introduced as a tool to calculate
critical exponents in three dimensional systems at a phase transition point~\cite{Wilson:1973jj}. As a rule, QFTs in
$d=4-2\epsilon$ possess nontrivial critical points with coupling constants being of order $\epsilon$. It allows one to calculate
critical dimensions as power series in $\epsilon$ and extrapolate results to $\epsilon=1/2$.
The current state of the art $\epsilon$-expansion technique and the corresponding references can be found, e.g. in
Refs.~\cite{Batkovich:2016jus,Kompaniets:2017yct}.

It is clear, however, that  QFTs in non-integer dimensions are not full fledged quantum field models~---~no
 real physical system is described by these QFTs.
 Thus they are not obliged to comply with expectations based on physical principles.  It was shown in
Ref.~\cite{Hogervorst:2015akt}, in the example of $\varphi^4$ theory, that such models are necessarily non-unitary. In the
$\varphi^4$ model, the lowest state with a negative norm is associated with an operator of rather high dimension ($\Delta=15$)
and the first complex anomalous dimensions appear for operators of dimension $\Delta=23$.
Therefore one may hope that the effect of these states to, e.g., the operator product expansion (OPE) could be neglected.
In the fermionic models, however,  the negative norm
operators have a rather low,
$\Delta=6$, canonical dimension~\cite{Ji:2018yaf} and can hardly be ignored.

Physical observables in conformal field theories (CFTs) are correlation functions of local operators. One is interested, in particular,
 in their
behavior under scale and conformal transformations. Therefore the basis of operators which transform in a proper way under
scale and conformal transformations plays a distinguished role. In perturbation theory, such a basis
is constructed by diagonalization of the anomalous dimension matrices. Since only
operators of the same canonical dimension mix under renormalization, such a matrix has a finite size in scalar field theories.
In a fermionic QFT, the situation is quite different~---~the number of mixing operators is, in most cases, infinite. The
simplest example of this kind is given by the four fermion operators,
\begin{align}\label{4q}
\mathcal{O}_n=\frac1{n!}\left(\bar q\, \Gamma_n^{\mu_1\ldots\mu_n} q\right)\left(\bar q \,\Gamma^n_{\mu_1\ldots\mu_n} q\right),
\end{align}
where $n=0,1,\ldots$ and $\Gamma^{n}_{\mu_1\ldots\mu_n}$ is the antisymmetrized product of the $d$-dimensional
$\gamma$-matrices. All these operators have canonical dimension $\Delta=6$ and mix under renormalization.
Customarily, the operators with $n\leq 4$ are called physical operators, and all others, since they vanish in $d=4$,
evanescent ones.

In the QCD context, four-fermion operators arise in the description of nonleptonic weak decays of  hadrons. Their renormalization
was studied in \cite{Buras:1989xd,Dugan:1990df,Herrlich:1994kh}. It was shown in~\cite{Dugan:1990df}  that the mixing between
evanescent operators and the physical ones can be avoided  by an appropriate  modification of the subtraction scheme.

Here we are interested in a different question --- constructing operators which have certain scaling dimensions at a
critical point. Since the size of the mixing matrix for the operators~\eqref{4q} is infinite, it is far from obvious that it
can be done in all situations. Solving the eigenvalue problem, one has to impose certain requirements (quantization conditions)
on the solutions.
Since we are interested in determining the scaling properties of the correlators of operators~\eqref{4q}, in particular the
simplest one
$\vev{\mathcal{O}_n(x) {\mathcal O}_{m}(0)}$,
it is reasonable to require  the correlation functions between two eigen-operators to be finite, i.e., for
$\mathcal{O}_{\Delta}=\sum_{n=0}^\infty
c_n(\Delta)\mathcal{O}_n$,
\begin{align}\label{quantization}
\langle\mathcal{O}_{\Delta}(x)\mathcal{O}_{\Delta}(0)\rangle <\infty.
\end{align}
This condition is always fulfilled if the mixing matrix has a finite size, as in the case of scalar field theories, but leads to
nontrivial ``quantization" conditions for infinite matrices.

In this work, we consider  renormalization of four-fermion operators in two theories: the critical QED and ${SU}(N_c)$
Gross-Neveu-Yukawa
(GNY) models. The first model (QED) was used as an example by  Dugan and Grinstein in their analysis~\cite{Dugan:1990df}. In
both cases, the spectral problem is equivalent to solving certain three-term recurrence relations. We present analytic
solutions to these recurrence relations and discuss the possibility to satisfy the condition~\eqref{quantization}.

The paper is organized as follows: In Sect.~\ref{sect:QED}, we present the solution to the one-loop mixing problem for the
operators~\eqref{4q} in QED. The operator mixing in the GNY model is discussed in Sect.~\ref{sect:GNY}. We introduce an
extended GNY model in Sect.~\ref{sect:GNYNc} and study the renormalization of four-fermion operators in this model.

\section{Critical QED}\label{sect:QED}

In $d=4-2\epsilon$ dimensions, QED with $N_f$ fermions has an infrared stable critical point at $a=a_*$ ($a=e^2/4\pi^2$)
\begin{align}
a_*=3\epsilon/N_f+O(\epsilon)\,.
\end{align}
At the critical coupling, the theory (in the Landau gauge)  is scale invariant~\footnote{ The model can also be analyzed with the
$1/N_f$ expansion technique, see Ref.~\cite{Gracey:1996tb} for a review and references. For a discussion of the three
dimensional model (QED$_3$) and its critical properties see, e.g.~\cite{Pisarski:1984dj,Appelquist:1988sr,Kotikov:2016wrb}. }.

Renormalization of four-fermion operators~\eqref{4q} in QED was studied in~\cite{Dugan:1990df}. In order to avoid
unnecessary complications, it is convenient to assume 
that the fermion and anti-fermion
fields have different flavors. At the critical point the renormalized operators $[{\mathcal O}]_n$ satisfy the renormalization group
equation,
\begin{align}
\Big(\delta_{nm}\,M\partial_M  +\gamma_{nm}\Big)[{\mathcal O}]_m=0\,,
\end{align}
where $M$ is the renormalization scale and $\gamma_{nm}$ is the anomalous dimension matrix. At one-loop, the matrix
$\boldsymbol{\gamma}$ takes the following form~\cite{Dugan:1990df}
\begin{align}\label{gamma_qed}
\gamma_{nm} &= \frac{a_*}2\big[(n+2)(n+1)\delta_{n+2,m}-2(n-1)(n-3)\delta_{n,m}
\notag\\
&\quad + (n-5)(n-6)\delta_{n-2,m}\big].
\end{align}
In order to construct an operator with a certain scaling dimension, $\mathcal{O}_\gamma=\sum_n c_n \mathcal{O}_n$, one has
to find the left eigen-vectors of the matrix~$\boldsymbol\gamma$,
\begin{align}\label{gqed}
\sum_n c_{n} \gamma_{nm}=\gamma\, c_m\equiv\frac12 a_*  \bar\gamma\, c_m\,.
\end{align}
%
Since there is no mixing between the operators $\mathcal{O}_n$ with odd and even index $n$, each set can be analyzed separately. The
analysis in both cases goes along the same line, and we therefore consider odd
$n$ only.

The transpose matrix $\gamma^T$ (from now on the indices $n,m$ take odd values) has a three-diagonal form
\begin{align}
\boldsymbol{\gamma}^T=\frac{a_*}2\begin{pmatrix}
a_1&b_3& 0& 0& 0& 0&\cdots
\\
d_1 & a_3& 0&0&0&0&\cdots
\\
0& d_3 & a_5 &b_7&0&0&\cdots\\
0&0 & d_5 &a_7&b_9&0&\cdots\\
\hdotsfor[1]{7}
\end{pmatrix},
\end{align}
where $a_n=-2(n-1)(n-3)$, $b_n=(n-5)(n-6)$, $d_n=(n+2)(n+1)$ and we take into account that $b_5=0$. The two-by-two block in the
upper--left corner describes mixing between the physical operators, $\mathcal{O}_1$ and $\mathcal{O}_3$. The corresponding
eigenvalues are
$\bar\gamma_\pm=\pm 6$. The eigenvectors corresponding to these eigenvalues take the form
\begin{align}
c^+_n=1,&& c^{-}_n=(n-2), &&n \text{ is odd}\,,
\end{align}
and the two operators $\mathcal{O}^\pm$ are
\begin{align}\label{Opm}
\mathcal{O}^+=\sum_{n\in \mathbb{N}_-} \mathcal{O}_n, &&\mathcal{O}^-=\sum_{n\in\mathbb{N}_-}(n-2) \mathcal{O}_n,
\end{align}
where sums go over odd integers.

All other eigenvectors of the matrix $\boldsymbol{\gamma}$  have the form $\vec{c}=(0,0,c_5,c_7,\ldots)$. Indeed, the subspace
spanned by these vectors is an invariant subspace of the matrix $\boldsymbol\gamma$. Looking for solutions in the form
$c_{2k+5}=s_{k}(2k+5)!/(2k)! $
with $k=0,1,2,\ldots$, one gets the following recurrence relation
\begin{align}\label{recrel}
C_k s_{k-1} -(A_k+C_k)s_k +A_k s_{k+1}=\frac14(\bar\gamma -26) s_k,
\end{align}
where $C_k=k(k-1/2)$ and $A_k=(k+3)(k+7/2)$. The above equation is nothing but the recurrence relation for the continuous
dual Hahn polynomials~\cite{MR2656096,MR1688958}. Its solutions take the form
\begin{align}\label{dualHahn}
s_k(\nu)={}_3F_2\left(\genfrac{}{}{0pt}{}{-k,3+i\nu, 3-i\nu}{3,7/2}\Big|1\right),
\end{align}
where $\nu$ is given by $\nu^2=(-\bar\gamma -10)/4$. For large $k$  the coefficients $s_k(\nu)$ have a power--like behavior
\begin{align}
s_k(\nu)\underset{k\to\infty}{\sim}  r(\nu) (2k)^{i \nu-3}+ r(-\nu) (2k)^{-i \nu-3} +\ldots\,,
\end{align}
where $r(\nu) =15\cdot  2^{i\nu}/\Gamma(3+i \nu)$. These functions form a complete  orthonormal system on
$L^2(\mathbb{R}_+)$~\cite{MR2656096,MR1688958}
\begin{align}\label{ss-int}
\int_0^\infty d\nu \mu(\nu)\,
s_k(\nu)\,s_n(\nu)
=\delta_{nk}\frac{(2k)!}{(2k+5)!}\,,
\end{align}
where
\begin{align}\label{measure-dual}
\mu(\nu)&=\frac1{225}\frac{\nu(1+\nu^2)(4+ \nu^2)}{\sinh\pi \nu}\,.
\end{align}
 In order to fix the allowed values of $\nu$, let us consider correlator of two eigen-operators
\begin{align}\label{Onu}
\mathcal{O}_\nu(x)=\sum_{k\geq 0} s_k(\nu) \mathcal{O}_{2k+5}(x)\,.
\end{align}
Note that the sum involves evanescent operators only. The operators $\mathcal{O}_\nu$, as follows from Eq.~\eqref{dualHahn}, are
even functions of
$\nu$,
$\mathcal{O}_\nu=\mathcal{O}_{-\nu}$.

The leading order correlator of two basic operators (all fields have different flavors)  was calculated in~\cite{Ji:2018yaf}
\begin{align}
\vev{{\cal O}_n(x) {\cal O}_m(0)} & =\delta_{mn}\frac{24\, \omega(n)}{\pi^8 |x|^{12-8\epsilon}}\,,
\end{align}
where
\begin{align}\label{wn}
\omega(n)=
\begin{cases}
 1/n!(4-n)! &
 n\leq 4\\
2\epsilon(-1)^n\, (n-5)!/n! &
n\geq 5
\end{cases}\,.
\end{align}
Note that for the evanescent operators, $n\geq 5$, the weight factor $\omega(n)$  is proportional to $\epsilon$ and sign
changing.

Then for the eigen-operators~\eqref{Onu}, one obtains
\begin{align}\label{OOcor}
\vev{\mathcal{O}_\nu(x)\mathcal{O}_{\nu'}(0)} &= |x|^{-12+8\epsilon} R(\nu,\nu')\,,
\end{align}
where the residue $R(\nu,\nu')$ is given by the sum
\begin{align}
R(\nu,\nu')\sim -\epsilon\sum_k \frac{(2k+5)!}{(2k)!} s_k(\nu) s_k(\nu')\,.
\end{align}
For  large $k$, the summand decays as  $k^{-1\pm i \nu \pm i\nu'}$ and $k^{-1\pm i \nu \mp i\nu'}$. Thus,  the sum diverges if
$\nu$ has a nonzero imaginary
part. For real
$\nu$, the correlator~\eqref{OOcor} can be understood in the sense of distributions.
Assuming that $\nu,\nu' \geq 0$ and taking into account Eq.~\eqref{ss-int}, we get
\begin{align}
\sum_k \frac{(2k+5)!}{(2k)!} s_k(\nu) s_k({\nu'})= \mu^{-1}(\nu)\delta(\nu-\nu').
\end{align}
For the correlator it results in
\begin{align}
\vev{\mathcal{O}_\nu(x)\mathcal{O}_{\nu'}(0)} &= -\frac{48\epsilon}{|x|^{2\Delta_\nu}}\mu^{-1}(\nu) \delta(\nu-\nu')\,,
\end{align}
where we have included the one-loop correction to the operator dimension $|x|^{-12+8\epsilon}\rightarrow|x|^{-2\Delta_\nu}$ with
\begin{align}
\Delta_\nu=6-4\epsilon-2a_*\left(\frac52+\nu^2\right).
\end{align}
Note that the anomalous dimensions of evanescent operators are negative.

The relation inverse to Eq.~\eqref{Onu} reads
\begin{align}
\mathcal{O}_{2k+5}(x)=\int_0^\infty d\nu \mu(\nu)\, s_k(\nu)\, \mathcal{O}_\nu(x).
\end{align}
It results in the following expression for the correlator of two (one-loop renormalized) evanescent operators~\eqref{4q}
\begin{align}\label{Onnprime}
\vev{\mathcal{O}_{n}(x)\mathcal{O}_{n'}(0)}=
-48\epsilon\int^\infty_0 d\nu \mu(\nu) \frac{ s_k(\nu) s_{k'}(\nu)}{|x|^{2\Delta_\nu}}\,.
\end{align}
where $n=2k+5,\ n'=2k'+5$.

\vskip 2mm

Coming back to the physical operators $\mathcal{O}_\pm$, we note that  these operators contain an infinite tail of
evanescent operators, see Eq.~\eqref{Opm}.   The contribution of the evanescent operators to the correlators,
$\vev{\mathcal{O}_\pm(x)\mathcal{O}_\pm(0)}$ is of order $\epsilon$ and, strictly speaking,  beyond our accuracy.
Nevertheless, we stress that the corresponding sum converges.

Since the operators $\mathcal{O}_\pm$ have different scaling dimensions, their correlator has to vanish. One can easily check
using~\eqref{Opm} that $\vev{\mathcal{O}_+(x)\mathcal{O}_-(0)}\sim O(\epsilon)$  as it should be. It can also be easily checked
that the correlator of an evanescent operator with the physical one is of order $\epsilon$ as well,
$\vev{\mathcal{O}_\nu(x) \mathcal{O}_\pm(0)}=O(\epsilon)$.


\section{Operator mixing in the Gross-Neveu-Yukawa model}\label{sect:GNY}
In this  section  we briefly consider the specifics of operator mixing in the GNY model~\cite{ZinnJustin:1991yn}.
The one-loop  anomalous dimension matrix for the four-quark operators~\eqref{4q} has been calculated in
Ref.~\cite{Ji:2018yaf}.
It has the following structure: the operator $\mathcal{O}_{n=0}$ is renormalized multiplicatively  and  the anomalous
dimensions matrix for the operators
$\mathcal{O}_n$, $n\geq1$  has a block-diagonal form
\begin{align}\label{diagA}
\boldsymbol{\gamma}= \mathrm{diag}(A_1,A_3,A_5,\ldots),
\end{align}
where each block $A_k$, with $k$ being odd, describes the mixing between the operators, $\mathcal{O}_k$ and $\mathcal{O}_{k+1}$.
The blocks
$A_k$  depend  nontrivially on $k$ but all have the same eigenvalues. Thus at the one loop-level, there are only two different
anomalous dimensions,
$\gamma_{\pm}$, which correspond to two different eigenvalues of the blocks $A_k$.
The anomalous dimension of the operator $\mathcal{O}_{n=0}$ coincides with $\gamma_+$.

Surprisingly enough, the  matrix $ \boldsymbol{\gamma}$ preserves this form at the two-loop order as well. We obtain the
following expression for the block $A_k$
\begin{align}
A_k &= 2u_*\left(1-u_*\frac{n_f+12}4\right) \begin{pmatrix}
k-1 & -1\\
-(k+1)(4-k) & 2-k
\end{pmatrix}
\notag\\
&\quad - \frac12 u_*^2 \begin{pmatrix}
19 & 0 \\
4(k+1)(4-n_f)& 4n_f+3
\end{pmatrix}\,,
\end{align}
where $n_f = N_f\times\tr\,\II$ and the critical value $u_*$ for the GNY model reads~\cite{ZinnJustin:1991yn,Mihaila:2017ble}
\begin{align}
u_*=\frac{2\epsilon}{n_f+6}\left(1+\frac{12\epsilon}{n_f+6}\right)+O(\epsilon^3)\,.
\end{align}
The eigenvalues of the blocks $A_k$ do not depend on $k$ 
\begin{align}\label{gamma_pm}
\gamma^+ &= 6 u_*\left(1-u_*\frac{7n_f+36}{12}\right)\,,
\notag\\
\gamma^- &= -4u_*\left(1-u_*\frac{2n_f+5}{8}\right)\,,
\end{align}
and the anomalous dimension of the operator $\mathcal{O}_{n=0}$ is still equal to $\gamma^+$.

Explanation of such degeneracy of the anomalous dimensions is the following: let us consider two sets of operators,
\begin{flalign}\label{OOnprime}
\mathcal{O}_n &=(\bar\psi_1 \Gamma_n\psi_2)(\bar \psi_3\Gamma_n\psi_4),
\notag\\
\mathcal{O}^\prime_n &=(\bar\psi_1 \Gamma_n\psi_4)(\bar \psi_3\Gamma_n\psi_2).
\end{flalign}
%
The  operator $\mathcal{O}_n$ and $\mathcal{O}^\prime_n$ obey exactly the same RG equation. At the same time they are related
to each other by Fierz transformation~\eqref{Fierz}
%
\begin{align}\label{Fierzrel}
\mathcal{O}^\prime_n=\sum_m\Omega_{nm}(d) \mathcal{O}_m\,.
\end{align}
Going over to the renormalized operators one gets
\begin{align}\label{O-Omega}
[\mathcal{O}^\prime]_n=\sum_m\widetilde \Omega_{nm}(d) [\mathcal{O}]_m\,,
\end{align}
where $[\mathcal{O}]_n = Z_{nm} \mathcal{O}_m$ ($[\mathcal{O}^\prime]_n = Z_{nm} \mathcal{O}^\prime_m$) and
\begin{align}
\widetilde \Omega(d) =Z \Omega(d) Z^{-1}\,.
\end{align}
The matrix $ \widetilde \Omega $ is a finite matrix (has no $\epsilon$ poles) of infinite size which depends on $\epsilon$ and the
coupling
constants. Taking the derivative
$M\partial_M$ on both sides of  Eq.~\eqref{O-Omega},
one finds that at the critical point, the anomalous dimension matrix $\boldsymbol{\gamma}$
commutes with $\widetilde \Omega$,
\begin{align}
\boldsymbol{\gamma}\,\widetilde \Omega=\widetilde \Omega\,\boldsymbol{\gamma}\,.
\end{align}
Then, provided that the matrix $\boldsymbol{\gamma}$ has a block diagonal form~\eqref{diagA}, it follows that the matrix
\begin{align}
\widetilde\Omega^{(km)}=\begin{pmatrix}
\widetilde \Omega_{k,m} & \widetilde \Omega_{k,m+1}\\
\widetilde \Omega_{k+1,m}&\widetilde \Omega_{k+1,m+1}
\end{pmatrix}
\end{align}
intertwines  the blocks $A_k$ and $A_m$
\begin{align}
A_k\, \widetilde \Omega^{(km)} = \widetilde \Omega^{(km)}\, A_m\,,
\end{align}
hence they have the same eigenvalues as $\widetilde\Omega^{(km)}$ is a convertible matrix.

In a similar manner, one can easily show that the vector
$\vec{c}_k =(\widetilde \Omega_{0,k},\widetilde \Omega_{0,k+1})$ is an
eigenvector of the matrix  $A_k^T$,
\begin{align}
A_k^T\,\vec{c}_k=\gamma_0 \,\vec{c}_k\,,
\end{align}
where $\gamma_0$ is the anomalous dimension of the operator~$\mathcal{O}_{n=0}$. Hence, $\gamma_0$ coincides with one of the
eigenvalues~\eqref{gamma_pm}, namely $\gamma_0=\gamma_+$.

Thus we conclude that as long as the matrix $\boldsymbol{\gamma}$ retains a  block-diagonal form, its eigenvalues will be
degenerate. We expect that the degeneracy of the anomalous dimensions in this model will be lifted by  the three--loop
corrections. It is, however,   simpler to consider a model where the degeneracy is absent already at the one-loop order.

\section{$SU(N_c)$ Gross-Neveu-Yukawa model}\label{sect:GNYNc}
We consider $SU(N_c)$ extension of the Gross-Neveu-Yukawa model~\cite{ZinnJustin:1991yn}. This model describes a system  of
interacting \underline{}fermion and scalar fields. (The bosonic model of this type was considered in
Ref.~\cite{Vicari:2006xr,Antonov:2013aka,Antonov:2017pqv}.) The fermion field has two isotopic indices, $q=q^{i,I}$ which
refer to the
$SU(N_c)$ and  $SU(N_f)$ global groups, respectively. The scalar field is in the adjoint representation of the $SU(N_c)$ group,
$\sigma=t^a \sigma^a$,
and we assume
%
the standard normalization $\tr t^a t^b=\frac12 \delta^{ab}$ for the generators $t^a$. The renormalized action takes the form
\begin{align}
{S}_R &=\int d^dx \Big(Z_1\bar q\slashed{\partial}q +{Z_2}\tr (\partial\sigma)^2
+ M^\epsilon Z_3 g\bar q \sigma q
\notag\\
&\quad + \frac1{4!}M^{2\epsilon}\Big(
Z_4\lambda_1 (\tr\sigma^2)^2 + Z_5\lambda_2 \tr\sigma^4
\Big)
\Big)\,.
\end{align}
For $N_c=2$ \ $(\tr\sigma^2)^2=2\tr\sigma^4$ so that one of the coupling becomes redundant and  can be put to zero, (we choose
$\lambda_1=0$). Introducing the notations
\begin{align}
 n_f=N_f\times \tr_\gamma \II,&& u=g^2/(4\pi)^2, && \bar\lambda_i=\lambda_i/(4\pi)^2\,,
\end{align}
one obtains the following  one-loop renormalization constants
\begin{flalign}
Z_1&=1-\frac u {2\epsilon}C_F&& Z_2=1-\frac {n_f u} {4\epsilon}, && Z_3=1-\frac{1}\epsilon \frac{u}{2N_c},
\end{flalign}
where $C_F=(N_c^2-1)/2N_c$ and
\begin{flalign}
Z_4&=1+\frac{\bar\lambda_1}{\epsilon}\frac{N_c^2+7}{24}+\frac{\bar\lambda_2}{6\epsilon}\left[N_c-\frac3{2N_c}\right]
+\frac1{8\epsilon}\frac{\bar\lambda_2^2}{\bar\lambda_1} \frac{N_c^2+3}{N_c^2},
\notag\\
Z_5&=1+\frac{\bar\lambda_2}{12\epsilon}\left[N_c-\frac{9}{N_c}\right]+\frac1{2\epsilon}
\bar\lambda_1-\frac{6}{\epsilon}\frac{n_f u^2}{\bar\lambda_2}.
\end{flalign}
For the index $\eta$ one gets
\begin{align}
\eta\equiv2\gamma_q=u C_F+O(\epsilon^2)\,.
\end{align}
The one--loop $\beta$ functions take the form
\allowdisplaybreaks{
\begin{flalign}
\beta_{\bar\lambda_1}&= \bar\lambda_1\Biggl(-2\epsilon  + n_f u  +\bar\lambda_1 \frac{N_c^2+7}{12}
   +  \bar\lambda_2 \frac{N_c^2-3}{6N_c} \Biggr)
\notag\\
&\quad + \frac1{4}\bar\lambda_2^2\left(1+\frac 3{N_c^2}\right),
\notag\\
\beta_{\bar\lambda_2}&=\bar\lambda_2\left(-2\epsilon  + n_f u +\bar\lambda_2 \frac{N_c^2-9} {6N_c} +\bar\lambda_1\right) -12 n_fu^2,
\notag\\
\beta_u&=2u\left(-\epsilon+u\left(\frac{n_f}4+\frac{N_c^2-3}{2N_c}\right)\right),
\end{flalign}
}
and for $N_c=2$  ($\lambda_1=0$)
\begin{align}
\beta_{\bar\lambda_2}&=\bar\lambda_2\left(-2\epsilon  + n_f u +\frac{11}{24}\bar\lambda_2\right)
-12 nu^2\,.
\end{align}
For the critical $u$-coupling one immediately gets
\begin{align}\label{u_crit}
u_*=4\epsilon/(n_f+2N_c-6/N_c)+O(\epsilon^2)\,.
\end{align}
To find the other two couplings we assume that $n_f\gg N_c$, Then one gets (up to $O(N_c/n_f)$ terms)
\begin{align}\label{lambda_crit}
\bar\lambda_2^*=\frac{96\epsilon}{n_f},
&& \bar\lambda_1^*=-\frac{1152\epsilon}{n_f^2}\left(1+\frac3{N_c^2}\right)\,.
\end{align}
The matrix $\omega_{ik}=\partial_{g_i} g_k$ at the critical point reads
\begin{align}
\omega=2 \epsilon \left(\II+O\left({ N_c }/{n_f}\right)\right)\,.
\end{align}
Since all eigenvalues of $\omega$ are positive, the critical point, $(u_*, \bar\lambda_1^*,\bar\lambda_2^*)$, is IR stable.
Note that although $\bar\lambda_1^*<0$, the scalar potential $V(\sigma)=\bar\lambda_1 (\tr\sigma^2)^2
+\bar\lambda_2\tr\sigma^4$ is positive since
$\bar\lambda_2^*+N_c\bar\lambda_1^*>0$.

Numerical analysis shows that the stable critical point exists for all  $N_c$ if $n_f$ is sufficiently large. For large
$N_c$, the necessary condition boils down to  $n_f>2N_c$.
 \vskip 2mm

Let us study the renormalization of four-fermion operators in this model. First, we note that the operators~\eqref{4q} are not
closed under renormalization and one has to consider the extended set of operators
\begin{align}
\mathcal{O}_n &=\frac1{n!}\left(\bar q\, \Gamma_n^{\mu_1\ldots\mu_n} q\right)\left(\bar q \,\Gamma^n_{\mu_1\ldots\mu_n} q\right),
\notag\\
\widehat{\mathcal{O}}_n &=\frac1{n!}\left(\bar q\, \Gamma_n^{\mu_1\ldots\mu_n} t^a q\right)\left(\bar q \,
t^a \Gamma^n_{\mu_1\ldots\mu_n} q\right).
\end{align}
Hereafter, we assume that all fields have different flavors. In order to write the anomalous dimension matrix, it is
convenient to organize the operators into the following multiplets,
\begin{align}
X_n^T=\big({\mathcal{O}}_n,
         \widehat{\mathcal{O}}_{n+1},
                    \widehat{\mathcal{O}}_{n+2},
                    {\mathcal{O}}_{n+3}\big)\,,
\end{align}
where $n=-1,1,3,\ldots$ (of course, the operator $\mathcal{O}_{n=-1}$  in  $X_{-1}$ has to be omitted.).

At the critical point the RGE for the operators $X_n$ can be written in the form
\begin{align}\label{XY}
\left(M\frac{d}{dM} + 2\eta+ H_n\right)X_n=- u_*\frac{N_c^2-4}{2N_c}\,Y_n,
\end{align}
where the matrix $H_n$ and vector $Y_n$ take the form
\begin{widetext}
\begin{align}
H_n=2u_*
\begin{pmatrix}
       C_F(2-n)                 &        n+1                     &  0               &   0 \\
  \frac{C_F}{2N_c}(4-n)    &  -\frac1{2N_c}(n-1)          &   -\frac{N_c}4(n+2)   & 0\\
     0                             &-\frac{N_c}4(3-n)     &  \frac1{2N_c}n    & \frac{C_F}{2N_c}(n+3)\\
    0                         &        0                       & 2-n       & C_F(n+1)
\end{pmatrix},
&&
Y_n=\begin{pmatrix}
0\\
(4-n) \widehat{\mathcal{O}}_{n}\\
(n+3)\widehat{\mathcal{O}}_{n+3}\\
0
\end{pmatrix}.
\end{align}

%
\end{widetext}
For $N_c=2$ the r.h.s. of Eq.~\eqref{XY} vanishes and  the anomalous dimension matrix acquires a  block-diagonal form,
with the block being equal to the matrix $H_n$. As could be expected from the discussion in the previous section the
eigenvalues of the block
$H_n$ do not depend on
$n$
and the anomalous dimensions take the following values
$$
\gamma=\left\{6u_*,\, \frac92u_*,\, 2u_*,\, -\frac32 u_*\right\}.
$$

For $N_c>2$ Eqs.~\eqref{XY} do not decouple  for different~$n$ and, although they can be reduced to the three term recurrence
relations, are still too complicated to be solved analytically.  The problem becomes more tractable in the large
$N_c$ limit. In this limit, $N_c\to\infty$, with $N_c/n_f$ being fixed, the operators with and without a hat
decouple from each other. Moreover, there is no mixing within the operators $\mathcal{O}_n$ themselves so that
each operator ${\cal O}_n$
evolves autonomously  in this limit. The anomalous dimensions of the operators $\mathcal{O}_n$ with even $n$ and odd $n$ are
\begin{flalign*}
\gamma^+_n=u_* N_c \, (n-1) +O(\epsilon^2),\! &&\!\gamma^-_n=u_*N_c (3-n)+O(\epsilon^2),
\end{flalign*}
respectively. At the same time, the operator $\widehat{\mathcal{O}}_n$ satisfies the following equation
%
\begin{flalign}
&\left(M\partial_M+2\eta\right)\widehat{\mathcal{O}}_n =\notag\\
&\qquad= \frac {u_* N_c}2 (-1)^n\big[(n+1)
\widehat{\mathcal{O}}_{n+1} +(n-5)\widehat{\mathcal{O}}_{n-1}
\big].
\end{flalign}
Looking for the eigen-operator in the  form
\begin{align}
\widehat{\mathcal{O}}=\sum_n (-1)^{\frac{n(n-1)}2} c_n \widehat{\mathcal{O}}_n,
\end{align}
%
one finds that, if the coefficients $c_n$ satisfy  the recurrence relation
\begin{align}\label{rec-sun}
2\lambda c_n=n \,c_{n-1}-(n-4)c_{n+1}\,,
\end{align}
then  $(M \partial_M+\gamma_\lambda) \widehat{\mathcal{O}}_\lambda=0$, where $\gamma_\lambda= u_* N_c(1-\lambda)$.

As it was discussed in Sect.~\ref{sect:QED}, the solutions to~\eqref{rec-sun} must ensure that the
correlator of eigen-operators  $\vev{\widehat{\mathcal{O}}_\lambda(x)\widehat{\mathcal{O}}_{\lambda'}(0)}$ is finite.

For the ``physical'' operators (such that not all $c_n=0$, for $n<5$), one easily obtains
\begin{flalign}
\widehat{\mathcal{O}}_{\lambda=\pm 2} & =\sum_n{(\pm 1)^n}{(-1)^{\frac12n(n-1)}} \widehat{\mathcal{O}}_n,
\notag\\
\widehat{\mathcal{O}}_{\lambda=\pm 1} & =\sum_n {(\pm 1)^n(n-2)}{(-1)^{\frac12n(n-1)}} \widehat{\mathcal{O}}_n,
\notag\\
\widehat{\mathcal{O}}_{\lambda=0} & =\sum_{n} {(-1)^{\frac12n(n-1)}}{(n-1)(n-3)} \widehat{\mathcal{O}}_n.
\end{flalign}
%

All other solutions of the recurrence relation~\eqref{rec-sun} have the following form
\begin{flalign}\label{sunck}
c_{k+5}(\lambda)& \equiv t_{k}(\lambda)
=\frac{1}{2\pi i} \oint \frac{dz}{z^{k+1}} (1-z)^{-3 + \lambda} (1+z)^{-3 - \lambda}\,
\notag\\
&=(-1)^k\frac{(3+\lambda)_k}{k!}
{}_2F_1\left(\genfrac{}{}{0pt}{}{-k,3-\lambda}{-k-2-\lambda}\Big|-1\right),
\end{flalign}
where the integration contour encircles the point $z=0$. Since the coefficients $c_n=0$ for $n<5$, the corresponding
eigen-operator is built from the evanescent operators only. The functions $t_k(\lambda)$ are polynomials  of degree
$k$ in
$\lambda$, (anti)symmetric under, $\lambda\to -\lambda$, $t_{k}(\lambda)=(-1)^{k} t_k(-\lambda)$.
The asymptotic of $t_k(\lambda)$ for large $k$ reads
\begin{flalign}
t_{k}(\lambda)&=\frac{k^{2-\lambda}}{\Gamma(2-\lambda)}
                +(-1)^{k}\frac{k^{2+\lambda}}{\Gamma(2+\lambda)}+\ldots.
\end{flalign}
They  form a complete orthonormal system
%
\begin{align}
\int_{-\infty}^{\infty} d\lambda\,\varkappa(\lambda)\, t_k(i\lambda)\overline{ t_j(i\lambda)}=\delta_{kj}\frac{(k+5)!}{32k!}
\end{align}
with respect to the measure
\begin{align}
\varkappa(\lambda)=\frac{\lambda(1+\lambda^2)(4+\lambda^2)}{ \sinh\pi\lambda},
\end{align}
which coincides, by a chance, with the measure~\eqref{measure-dual}. It implies, in particular, that
%
$t_{2k} (i\lambda)\sim s_k(\lambda).$
%
We discuss it in more details in Appendix~\ref{app:Hahn}.

 In order to fix the allowed values of $\lambda$, we consider the  correlator of two  eigenoperators. At the leading order it
takes the form
\begin{align}\label{cor-lambda}
\vev{\widehat{\mathcal{O}}_\lambda(x)\widehat{\mathcal{O}}_{\lambda'}(0)} \sim  |x|^{-12+8\epsilon} R(\lambda,\lambda')\,,
\end{align}
where the residue $R$ is given by the sum, (see Eq.~\eqref{wn})
\begin{align}
R(\lambda,\lambda') = \sum_{k>0} (-1)^k \frac{k!}{(k+5)!} t_k(\lambda) t_k(\lambda')\,.
\end{align}
The sum diverges unless $\text{Re} \lambda=0$. For imaginary $\lambda$ the correlator~\eqref{cor-lambda} exists in the sense of
 distributions. Thus the anomalous dimensions of the operator $\widehat{\mathcal{O}}_\lambda$ is complex,
$\gamma_\lambda=u_*N_c(1-\lambda)$.

\vskip 2mm

\vskip 2mm

One notices that there is a certain resemblance  between  the anomalous dimensions of four-fermion operators in the
$SU(N_c)\times SU(N_f)$ GNY
model and QED. 
Mixing among evanescent operators results in a continuous spectrum. In QED, the anomalous dimensions stay real,
although negative, while in the GNY model they become complex. Of course, it is not excluded that this effect is an artifact
of the one-loop approximation. Indeed, the spectrum is mainly determined by  details of the anomalous dimension matrix at large
$n$.
At one loop,  the matrix elements
$\gamma_{nm}$ grow with $n$ as
$\epsilon n^2$  and $\epsilon n$ in QED and in the GNY model, respectively.
One has all reasons to expect that higher-order corrections will scale as
$(\epsilon n^2)^k$ and $(\epsilon n)^k$. Whenever $\epsilon n, \epsilon n^2 \sim O(1)$, these corrections have to  be
resummed. Such a resummation can drastically  change the large $n$ behavior of the matrix elements~~\footnote{In QED the anomalous
dimension matrix in the physical sector at two loops were obtained in~\cite{DiPietro:2017kcd}.}.

\vskip 5mm
 Finally, we consider an example to show that the construction of  operators with ``good" scaling properties is not always
possible. Let $\mathcal{O}_n$ and $\mathcal{O}^\prime_n$ be the operators introduced in the section \ref{sect:GNY},
Eq.~\eqref{OOnprime}. These two sets of operators are related to each other by the Fierz transformation~\eqref{Fierzrel}. The
correlators of the operators
\begin{flalign}
f_{nm}(x) & =\vev{\mathcal{O}_n(x) \mathcal{O}_m(0)}=\vev{\mathcal{O}^\prime_n(x) \mathcal{O}^\prime_m(0)}\,,
\notag\\
f^\prime_{nm}(x) & =\vev{\mathcal{O}_n(x) \mathcal{O}^\prime_m(0)}\,,
\end{flalign}
are well defined in the perturbative expansion (here, for  definiteness, we consider QED model) and satisfy the same RGEs.
Namely, for $\varphi = f, f^\prime$ one gets (at the critical point)
\begin{align}
M\partial_M \varphi_{nm}+ (\gamma \varphi)_{nm} +(\varphi \gamma^T)_{nm}=0\,.
\end{align}
Going over to the operators $\mathcal{O}_\nu=\sum_n c_n(\nu) \mathcal{O}_n$ ($\mathcal{O}'_\nu=\sum_n c_n(\nu)
\mathcal{O}^\prime_n)$ one can bring the correlator $f_{nm}$ into the form~\eqref{Onnprime}. The coefficients $c_n(\nu)$ are
determined by two conditions: first, they have to diagonalize the matrix $\gamma$, $\gamma_{nm} c_m(\nu) \sim c_{n}(\nu)$ and
second, the product $(c_n(\nu) f_{nm}(x) c_m(\nu'))$ should exist in the sense of distributions.

Proceeding along the same lines with the correlator $f^\prime_{nm}$ one finds that while the first condition leads to the
same vectors $c_n(\nu)$, the normalization condition changes. Now it reads (at the leading order in $\epsilon$)
\begin{align}\label{Rprime}
R'(\nu,\nu')\sim \sum_{nm}c_n(\nu) \Omega_{nm} \omega_m c_m(\nu') <\infty\, .
\end{align}
The matrix $C_{nm}=\Omega_{nm} \omega_m$ is symmetric in $n,m$ and grows as $\sim n^{m}$ for large $n$ and fixed
$m$. It is easy to see that the sums in Eq.~\eqref{Rprime} diverge for any $\nu,\nu'$. It means that while $f'_{nm}$
correlators satisfy exactly the same RGE as $f_{nm}$, the former cannot be brought to the form~\eqref{Onnprime}.

This statement can also be formulated as follows. The matrix $\gamma$ commutes with the matrix $\Omega$, e.g., $\gamma \Omega=\Omega
\gamma$. However, while $c_n(\nu)$ is an eigenvector of $\gamma$, it does not belong to the domain of the operator $\Omega$,
i.e., $\sum \Omega_{nm} c_m(\nu)$ diverges.

The conclusion is that in  non-integer dimensions, the possibility of representing the correlator
$\vev{\prod_k \mathcal{O}_k(x_k)}$ as a sum
of the correlators with ``good" scaling properties depends on the operators $\mathcal{O}_k$ in question.

\section{Summary}
We have considered the renormalization of four--fermion operators in the critical QED and extended GNY models. The anomalous
dimension matrix in both models is of infinite size so that in order to make the diagonalization problem well defined,  additional
restrictions have to be imposed on the solutions. It is natural to demand for the correlation functions of the eigenoperators to be
finite in the $\epsilon$--expansion, Eq.~\eqref{quantization}. By diagonalizing the anomalous dimension matrix in both models, we
found that in both cases the spectrum is continuous and, for the extended GNY model, complex. Moreover, we argued that not all
correlators can be expanded as a sum (integral) of contributions with specific scale dependence. We expect that all these
properties are likely to be true in general for theories with fermions in $d<4$ dimensions.

It is expected that in  the $d\to 3$ limit, these continuous spectrum operators  should somehow decouple from the physical
operators so that the evanescent operators can be consistently put to zero. Clearly, this property is hard to check within the $\epsilon$-expansion  
where only a few terms in the series could be calculated. It seems that alternative approaches such as
the $1/N$ expansion are better suited for this purpose.

\acknowledgments
We are grateful to Michael Kelly for collaboration in the early stages of this project and V. Braun for useful comments.
This work  was supported by the DFG grants BR 2021/7-1 (YJ), MO~1801/1-3 (AM) and by RSF 
project 
14-11-00598 (AM).
\appendix
\section{ $\gamma$ matrices in $d$-dimensions}\label{app:Fierz}
The antisymmetrized product of $ \gamma$ matrices is defined as
\begin{align}
\Gamma^{(n)}_{\boldsymbol{\mu}}\equiv\Gamma^{(n)}_{\mu_1\ldots\mu_n}=
    \frac1{n!} \sum_P (-1)^P \gamma_{\mu_{i_1}}\ldots\gamma_{\mu_{i_n}}\,,
\end{align}
where the sum is taken over all permutations. Below we collect some formulas which were helpful for the calculation. The effective
technique for handling $\gamma$-matrices  can be found in Refs.\cite{Kennedy:1981kp,Vasiliev:1995qj}. Let us denote
\begin{align}
\Gamma_n\otimes \Gamma_n\equiv \Gamma^{(n)}_{\mu_1\ldots\mu_n}\otimes \Gamma^{(n)}_{\mu_1\ldots\mu_n}\, .
\end{align}
Then one  finds
\begin{align}
\gamma^\mu\Gamma_n \gamma_\mu \otimes \Gamma_n &=(-1)^n (d-2n)\Gamma_n\otimes \Gamma_n
\notag\\
\gamma_\mu \Gamma_n\otimes \gamma^\mu \Gamma_n &=\Gamma_n\gamma_\mu \otimes \Gamma_n\gamma^\mu
=\Gamma_{n+1}\otimes \Gamma_{n+1}
\notag\\
&\quad + n(d-n+1)\Gamma_{n-1}\otimes \Gamma_{n-1}
\notag\\
\gamma_\mu \Gamma_n\otimes \Gamma_n\gamma^\mu &=\Gamma_n\gamma_\mu \otimes \gamma^\mu \Gamma_n
=
(-1)^n \Big(\Gamma_{n+1)}\otimes \Gamma_{n+1}
\notag\\
&\quad - n(d-n+1)\Gamma_{n-1}\otimes \Gamma_{n-1}\Big).
\end{align}
The Fierz identity in $d$-dimensions has the form
\begin{align}\label{Fierz}
\Gamma_{n}^{\alpha\beta}\otimes \Gamma_{n}^{\gamma\delta}=\sum_{m=0}^\infty \Omega_{nk}(d)\,
\Gamma_{m}^{\alpha\delta}\otimes \Gamma_{m}^{\gamma\beta}\,.
\end{align}
The  Fierz coefficients $\Omega_{nm}$  can be written as follows~\cite{Vasiliev:1995qj}
\begin{align}\label{OmegaF}
\Omega_{nm}(d)&=\frac12 (-1)^{\frac{n(n-1)}2+\frac{m(m-1)}2} /\tr\II\notag\\
&\quad\times\big[1+(-1)^m+(-1)^n-(-1)^{n+m}\big]
\notag\\
&\quad\times \frac1{m!}\frac{d^n}{dx^n}(1+x)^{d-m}(1-x)^m\Big|_{x=0}\,.
\end{align}
The matrix $\Omega$ has to satisfy the consistency relation,
\begin{align}\label{consistency}
\sum_{m=0}^\infty \Omega_{nm}(d)\Omega_{mk}(d)=\delta_{nk}\,.
\end{align}
The series converges in the region $n,k\leq d$ and for other $d$, it has to be understood as an analytical
continuation~\cite{Avdeev:1983yg}. With the help  of the representation~\eqref{OmegaF}, the sum in~\eqref{consistency} can be
easily evaluated resulting in
$\delta_{nk}\,2^{d}/\tr^2_\gamma \II $. Thus the consistency relation
\eqref{consistency} holds only if the trace of the unit matrix is chosen to be~\cite{Avdeev:1983yg},
\begin{align}
\tr_\gamma \II =2^{d/2}.
\end{align}
Notice that this expression coincides  with the dimensions of the canonical (finite-dimensional)
$ \gamma$ matrix representation only for {\it even} $d$.

\section{ Hahn polynomials}\label{app:Hahn}
Here we collected some basic facts about the dual continuous Hahn polynomials, $S_n(x^2, a,b,c)$, which are defined
as~\cite{MR2656096}
\begin{align}
S_n(x, a,b,c)={}_3F_2\left(\genfrac{}{}{0pt}{}{-n,a+ix, a-ix}{a+b,a+c}\Big|1\right)\,.
\end{align}
They satisfy the recurrence relation
\begin{flalign*}
(A_n+C_n-a^2-x^2)S_n(x) &=C_n S_{n-1}(x)
+ A_n S_{n+1}(x),
\end{flalign*}
where
\begin{flalign*}
C_n =n(n+b+c-1)\,,
&&
 A_n =(n+a+b)(n+a+c).
\end{flalign*}
They form a complete orthonormal system on $\mathbf{L}^2(\mathbb{R}_+)$,
\begin{multline}\label{ort}
\frac1{2\pi}\int_0^\infty dx\, w(x,a,b,c) S_n(x) S_m(x)
=
\\=\delta_{mn}n!\Gamma(n+b+c)\frac{\Gamma(a+b)\Gamma(a+c)}{(a+b)_n(a+c)_n}\,,
\end{multline}
where the measure function reads
\begin{align}
w(x,a,b,c)=\frac{|\Gamma(a+ix)\Gamma(b+ix)\Gamma(c+ix)|^2}{|\Gamma(2ix)|^2}.
\end{align}
Next, we demonstrate that the polynomials in Eq.~\eqref{dualHahn} and \eqref{sunck} coincide, $s_n(\lambda) \sim
t_{2n}(i\lambda)$.
%
Let us consider the recurrence relation
\begin{align}
2\lambda p_n=(n+2\mu-1)p_{n-1}-(n+1)p_{n+1}\, .
\end{align}
which for $\mu=3$ is the recurrence relation for the polynomial $t_n(\lambda)$. The solutions have the form
\begin{flalign}
p_{n}(\lambda)
& =\frac{1}{2\pi i} \oint \frac{dz}{z^{n+1}} {(1-z)^{-\mu + \lambda}}{(1+z)^{-\mu - \lambda}}\notag\\
&=(-1)^n\frac{(\mu+\lambda)_n}{n!}
{}_2F_1\left(\genfrac{}{}{0pt}{}{-n,\mu -\lambda}{1-n-\mu-\lambda}\Big|-\!1\right).&
\end{flalign}
%
After rescaling $\displaystyle t_n={(2\mu)_n}/{n!}b_n$ the recurrence relation takes the form
\begin{align}\label{bkrec}
2\lambda b_n=n b_{n-1}-(n+2\mu)b_{n+1}\,.
\end{align}
After some algebra it can be transformed to the equation
\begin{flalign}\label{newrel}
4(\lambda^2-\mu^2) b_n &=(n+2\mu)(n+2\mu+1)(b_{n+2}-b_n)
\notag\\
&\quad +n(n-1) (b_{n-2}-b_n)
\end{flalign}
which involves the even/odd polynomials $p_n$ only. Having put $n=2k$ ($n=2k+1$) one find that \eqref{newrel} coincides with
the defining relation for the continuous dual Hahn polynomials, $S_k( i\lambda,\mu, 0,1/2)$ for even $n$, and
$S_k( i\lambda,\mu, 1/2,1)$ for odd one.
Taking into account the initial conditions $p_{n=0}=b_0=1$ ($b_1=-\lambda/\mu$) one gets
\begin{multline}\label{1rel}
{}_3F_2\left(\genfrac{}{}{0pt}{}{-k,\mu+\lambda, \mu-\lambda}{\mu,\mu+\frac12}\Big|1\right)=
\\
=\frac{(\mu+\lambda)_{2k}}{(2\mu)_{2k}}{}_2F_1\left(\genfrac{}{}{0pt}{}{-2k,\mu-\lambda}{1-2k-\mu-\lambda}\Big|-1\right)
\end{multline}
and
\begin{multline}
\frac{\lambda}{\mu}{}_3F_2\left(\genfrac{}{}{0pt}{}{-k,\mu+\lambda,\mu-\lambda}{\mu+\frac12,\mu+1}\Big|1\right)=
\\
=\frac{(\mu+\lambda)_{2k+1}}{(2\mu)_{2k+1}}{}_2F_1\left(\genfrac{}{}{0pt}{}{-2k-1,\mu-\lambda}{-2k-\mu-\lambda}\Big|-1\right).
\end{multline}
Having put $\mu=3$ in the relation~\eqref{1rel} one finds that $s_k(\lambda)= 6(2k)!/(2k+5)! t_{2k}(i\lambda)$.

\bibliography{ref_gny}
\end{document}